\documentclass[useAMS,usenatbib]{mn2e}

\usepackage{graphicx}
\usepackage{amssymb}
\usepackage{cite}

\newcommand{\dd}{{\rm d}} 
\newcommand{\Umag}{U_{\rm mag}} 
\newcommand{\Urad}{U_{\rm rad}} 

\newcommand{\gam}{\left({\gamma\over 10^4}\right)}
\newcommand{\bg}{\left({B\over 1\,{\rm G}}\right)}
\newcommand{\vsc}{\left({v_s\over c}\right)}
\newcommand{\Gam}{\left({\Gamma\over 50}\right)}
\newcommand{\tv}{\left({t_{\rm var}\over 300\,{\rm s}}\right)}
\newcommand{\den}{\left({n\over 10^5\,{\rm cm}^{-3}}\right)}
\newcommand{\blob}{\left({R\over 10^{15}\,{\rm cm}}\right)}

\title[Particle-acceleration timescales]%
      {Particle-acceleration timescales in TeV blazar flares}

\author[]{%
  Joni Tammi and Peter Duffy \\
  UCD School of Physics, University College Dublin, Belfield, Dublin 4, Ireland
}
\begin{document}

\date{Accepted 2008 November 18. Received 2008 November 14; in original form 2008 October 17.}

\pagerange{\pageref{firstpage}--\pageref{lastpage}} \pubyear{2008}

\maketitle

\label{firstpage}

\begin{abstract}
  Observations of minute-scale flares in TeV Blazars place constraints
  on particle acceleration mechanisms in those objects. The
  implications for a variety of radiation mechanisms have been
  addressed in the literature; in this paper we compare four different
  acceleration mechanisms: diffusive shock acceleration, second-order
  Fermi, shear acceleration and the converter mechanism. When the
  acceleration timescales and radiative losses are taken into account,
  we can exclude shear acceleration and the neutron-based converted
  mechanism as possible acceleration processes in these systems. The
  first-order Fermi process and the converter mechanism working via
  SSC photons are still practically instantaneous, however, provided
  sufficient turbulence is generated on the timescale of seconds. We
  propose stochastic acceleration as a promising candidate for
  the energy-dependent time delays in recent gamma-ray flares of
  Markarian 501.
\end{abstract}


%
\section{Introduction}

Rapid TeV variability has recently been observed by the HESS and MAGIC
telescopes in two blazars; PKS 2155--304 \citep{AharonianEtAl2007} and
Markarian 501 \citep{AlbertEtAl2007}. The timescales of a few minutes are
shorter, by at least an order of magnitude, than the light crossing
time of the central black hole with a mass of order $M=10^9\,{\rm
  M}_\odot$.  This suggests that the variability is associated with
small regions of the highly relativistic jet rather than the central
region, although the latter possibility cannot be completely excluded.
\citet{BegelmanEtAl2008} have examined the constraints that
these observations place on the size and location of the emitting
region. With an observed variability timescale $t_{\rm var}$ and a jet
Lorentz factor $\Gamma$, the flare, if it moves out with the flow,
occurs at a distance greater than $c t_{\rm var}\Gamma^2$.  For
$\Gamma\sim 50$ \citep{BegelmanEtAl2008} and minute timescale
variability this places the flaring region at a distance in excess of
one hundred Schwarzschild radii ($r_S$) from the central black hole.
The relativistic particles that are ultimately responsible for the
emission are then required to have been ejected from the central
region along with the jet, and subsequently survived out to $100 \, r_S$
where they then radiate away their energy quickly, or alternatively
the particles are accelerated within the jet itself, close to the
emission region. In this paper we examine the latter possibility and
compare the predicted acceleration timescales for four acceleration
mechanisms with the requirements for producing a minute timescale
flare.

There are three fundamental questions that we address in this paper.
Firstly, we want to see whether particle acceleration could be
considered instantaneous on the flaring timescale; is it realistic to
assume a fully-developed power-law injection spectrum at the onset of
flaring?  Related to this question is whether we can exclude
particular acceleration mechanisms on the grounds that they are too
slow.  Secondly, we are interested to see if we can find a set of
parameters for which only cooling would be present. If we assume
a given input spectrum for particles in a blob travelling with the
jet, can we expect further acceleration to be always present?
Alternatively, is it more likely that within minute to hour
timescales particles would only lose their energy instead of being
able to continue accelerating on the same timescales?  Finally, can
any of the mechanisms work on timescales comparable to the lags
between lower- and higher-energy radiation observed in some objects,
and what mechanisms could explain the acceleration of particles
on the timescale of minutes?

In Section 2 we discuss four possible acceleration mechanisms and
their associated timescales. These are (i) first-order Fermi
acceleration, (ii) second-order Fermi, (iii) acceleration in a shear
flow and (iv) the converter mechanism. The role played by
radiation mechanism and loss timescales are then compared with the
acceleration properties in Section 3.

%
\section{Acceleration timescales} \label{sec:t_acc}

When particles are scattered by magnetic fluctuations,
they gain energy whenever two subsequent scattering centres are moving
toward each other, leading to a ``head-on'' collision. Suitable
conditions are provided around a shock wave, where a relativistic
particle crossing the shock always sees the plasma -- and thus the
scattering centres -- of the flow on the other side of the shock,
approaching. Similarly, relative motion of the scattering centres
(Alfv\'en waves, for example) inside a constant-velocity plasma,
moving in different directions, are more often seen as approaching
than receding, leading to average gain in energy. The
average energy gain during a crossing cycle, and the duration of such
a cycle, determines the energy gain rate, $< \dd \gamma / \dd
t>$, and the acceleration timescale
\( \tau \equiv \gamma / \left< \dd \gamma / \dd t \right> \).

The timescale determines how long it takes for a particle with Lorentz
factor $\gamma$ to gain or lose energy, allowing easy comparison
between the efficiencies of different gain and loss mechanisms.  For
easily comparable and general results we assume the acceleration to
take place in a region with a characteristic, comoving spatial scale
$R$, with plasma flow having Lorentz factor $\Gamma$. Particles are
assumed to be scattered by irregularities in large-scale magnetic
field (with strength $B$) with the average free path between
scattering, $\lambda$, being equal to the particle gyroradius $r_g$.

%
\subsection{First-order Fermi acceleration}

The first-order process takes place at the shock front, where the
accelerating particles gain energy by crossing and re-crossing the
shock. An average cycle increases the
particle energy by a factor of $\Gamma^2$ for the first cycle, and
by a factor of $\sim 2$ thereafter. The duration of such
a cycle, as well as the probability for a particle to be injected into
one, depends heavily on the details of the scattering of the particles
in the turbulent plasma and the geometry of the shock. 
 In the Bohm limit where the particle's mean free path is equal to 
its gyroradius, $\delta B\sim B$, in which case
\begin{equation}\label{eq:r_g}
  r_{\rm g} = \frac{\gamma m c^2}{eB}
  \approx 1700 \, \sqrt{\gamma^2-1}\bg^{-1} \; {\rm cm}.
\end{equation}
In this case the acceleration timescale can be simplified to
\begin{equation}\label{eq:t_fermi1}
  \tau_{\rm FI} \gtrsim
  6 \, \left( \frac{c}{v_{\rm s}} \right)^2 \frac{\lambda}{c} \approx
  6 \, \frac{r_g c}{v_{\rm s}^2},
\end{equation}
where $v_{\rm s}$ is the speed of the shock. For a one-Gauss magnetic
field and a relativistic shock ($v_{\rm s} \to c$) this gives, for an
electron with $\gamma = 10^4$, acceleration timescale of a few
milliseconds.
For lower energy particles the acceleration is even
faster. With time resolution of the observations of the
order of minutes, this is ``instantaneous'' and poses no problems in
getting the particles to sufficiently high energies. Requiring the
acceleration rate to be less than variability timescale $t_{\rm var}$
(in the lab frame), requires that
\begin{equation}\label{eq:fermi1constraint}
  \gam\bg^{-1}\vsc^{-2}\Gam^{-1} < 4.4\times 10^6\,\tv
\end{equation}

For protons, however, whose mass and acceleration timescale are one
thousand times larger than for electrons, acceleration can take
several minutes.  When the observed timescales are of the order of
minutes, this leads to a situation where, for hadronic models, the
particle injection spectrum develops during the flare and has a
gradually increasing maximum energy cutoff even if the observed
timescales are reduced by strong Doppler boosting. For the remainder
of this paper we will concentrate on electron
acceleration. Furthermore, a source of radius $R$ cannot confine
particles with gyro-radius larger than $R$, so that there is a
geometric constraint, $r_g<R$, on the maximum particle
energy. However, this does not place an important constraint on the
upper-cutoff of the particle distribution for these sources.

%
\subsection{Converter mechanism}

In the converter process the accelerating particles cross a shock from
the downstream to the upstream in a neutral form, as a neutron or
synchrotron photon. This neutral particle then decays into a proton
and an electron (in the case of a neutron) or produces an
electron-positron pair (for a photon), which can then be scattered
again across the shock, and the process continues as in the
first-order mechanism until the accelerating particle enters
into the upstream region again in a neutral form. In the converter
mechanism, the average angle at which the upstream particle re-enters
the shock is larger than for the first-order process, and in the
converter process a particle can get the maximal $\Gamma^2$ boost in
every cycle instead of just the first one, as was the case for the
first-order acceleration \citep{DerishevEtAl2003,Stern2003}.

The scenario involving the shock-crossing in forms of free neutrons
can probably be excluded from the dominating process in this case, as
the free neutron lifetime of $\sim 15$ minutes, makes the whole cycle
too long for these minute-scale flares even for $\Gamma=50$
(corresponding to co-moving timescales of a few hours), because
multiple shock-crossing cycles are needed for the particles to reach
sufficient energies to account for all of the radiation. For
longer duration flares, however, even the neutron-based mechanism can
be plausible.

Instead, we concentrate on a variant of this process where synchrotron
photons play the role of the neutral particle. This process is
expected to work best in ultrarelativistic shocks where it competes
with the first-order mechanism. The converter mechanism is slower at
lower energies, but soon reaches the rate of the first-order process
-- it can also reach higher energies than the first-order one
\citep{DerishevEtAl2003}, and thus dominate the acceleration of the
highest-energy particles.  Because both the converter mechanism
working via synchtrotron photons and the first-order process in
general work on similar timescales, we approximate the converter
mechanism to work on timescale corresponding to that of the
first-order mechanism and incorporate its effect in the first-order
electron acceleration timescale calculations.

%
\subsection{Second-order Fermi  acceleration}

The second-order, or stochastic, process, 
accelerates particles using scattering centres moving
relative to each other even without differences in the actual flow
speed. For example, Alfv\'en waves in the turbulent downstream of a
relativistic low-Mach-number shock can provide promising conditions
for efficient stochastic acceleration \citep[]{VV2005ApJ} with
and without a shock. Because the process is not tied to the plasma speed,
it can continue to accelerate particles far away from the shock and
 for much longer than the first-order process -- provided there is
sufficient turbulence present.

The acceleration timescale for stochastic acceleration 
is \citep[]{RiegerEtAl2007}
\begin{equation}\label{eq:t_fermi2}
  \tau_{\rm FII} \approx 
  \frac{3}{4} \, \left( \frac{c}{v_{\rm A}} \right)^2 \, \frac{\lambda}{c}
  \approx \frac{3}{4} \, \frac{c r_g}{v_{\rm A}^2},
\end{equation}
where the Alfv\'en speed, defined by
\begin{equation} \label{eq:v_a}
  v_{\rm A}^2 = \frac{(B \, c)^2 }{4 \, \pi \, h\, n + B^2},
\end{equation}
depends on the enthalpy, \( h = (\rho + P)/n \), with the
energy density of the plasma, \( \rho = n \, m \, c^2 \), being a
function of the composition and number density , $n$. The
mass $m$ depends on the composition and is $m_{\rm ee}
= 2 \, m_{\rm e}$ for pure electron--positron plasma, and $m_{\rm ep} =
m_{\rm e} + m_{\rm p}$ for ionised hydrogen. The effect of the
gas pressure, $P$, is taken to be negligible.

The second order process will be rapid enough to occur on a timescale 
shorter than the observed flaring time provided that 
\begin{eqnarray}\label{eq:stochconstraint1}
  \gam\bg^{-3}\Gam^{-1}
  \left[a\den+\bg^2\right]\nonumber\\
  <3.5\times 10^6\tv
\end{eqnarray}
where $a = 2.1$ for an electron-positron plasma and $a=1.9\times
10^{3}$ for the hydrogen case. In a ``highly magnetised plasma''
\begin{equation}\label{eq:highmag}
  a\den\ll\bg^2
\end{equation}
the above constraint for a rapid stochastic process simplifies to 
\begin{equation}\label{eq:stochconstraint2}
  \gam\bg^{-1}\Gam^{-1} < 3.5\times 10^6\tv .
\end{equation}

\begin{figure}
  \begin{center}
    \includegraphics[width=\linewidth]{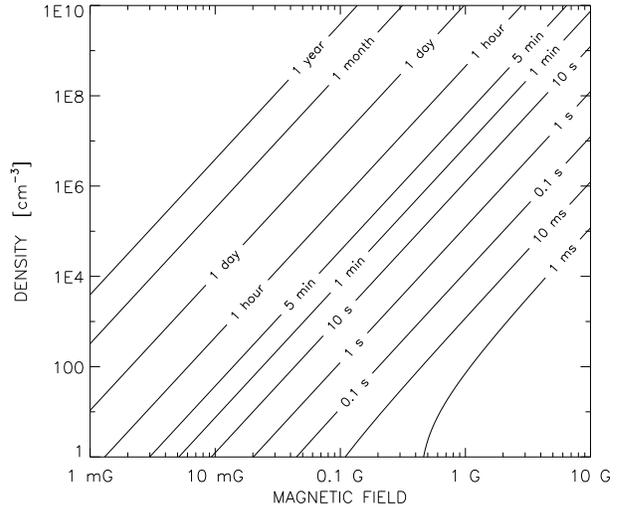}
  \end{center}
  \caption{Stochastic-acceleration timescale $\tau_{\rm FII}$
    (Eq.~\ref{eq:t_fermi2}) for an electron with $\gamma = 10^4$
    as a function of plasma number density $n$ and the magnetic field
    intensity $B$ in ionised hydrogen plasma. Timescales vary from
    years (top left corner) to milliseconds (bottom right).}
  \label{fig:t_fermi2_pe}
\end{figure}

\begin{figure}
  \begin{center}
    \includegraphics[width=\linewidth]{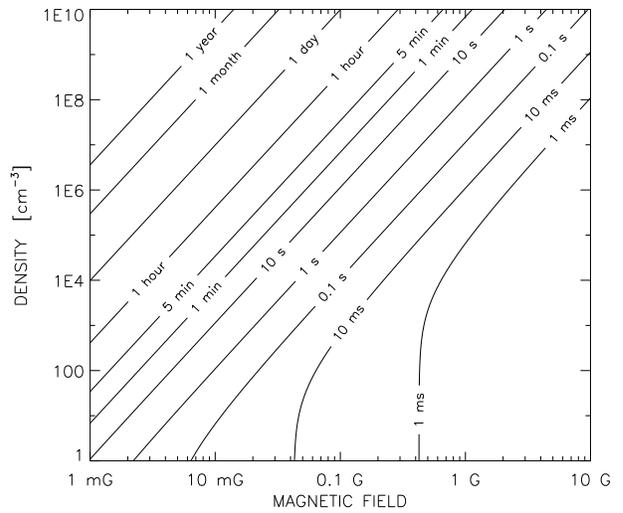}
  \end{center}
  \caption{As Fig.~\ref{fig:t_fermi2_pe}, but for pure pair plasma.}
  \label{fig:t_fermi2_ee}
\end{figure}

\begin{figure}
  \begin{center}
    \includegraphics[width=\linewidth]{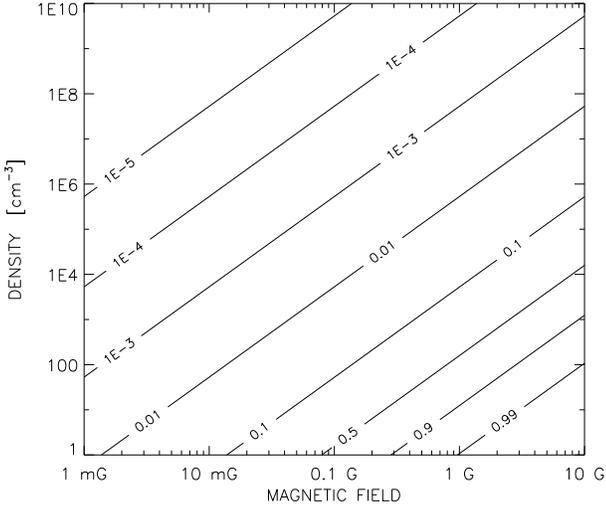}
  \end{center}
  \caption{Alfv\'en speeds, in the units of $c$, for parameters
    corresponding to those in Fig.~\ref{fig:t_fermi2_pe}.}
  \label{fig:alfven_pe}
\end{figure}

The stochastic acceleration timescale as a function of the magnetic
field and the number density of the plasma is shown in
Fig.~\ref{fig:t_fermi2_pe} for hydrogen plasma and in
Fig.~\ref{fig:t_fermi2_ee} for a pair plasma. Although the
acceleration is very slow when the magnetic field is relatively low
and the density is high, sites such as magnetically dominated AGN jets
with relatively low matter density and compressed magnetic fields
could favour fast acceleration with the blob matter density of the
order of $\sim 10^3$--$10^6$ particles (protons and electrons) per
cm$^3$ and a magnetic field of the order of one gauss, thus
providing acceleration timescales comparable to the observed
minute-scale flickering. For purely hadronless pair plasma the
acceleration timescale is much shorter, and can even be
``instantaneous'' if the plasma density is not very high. In sources
where the plasma is purely or mainly leptonic and has low density,
sufficiently high magnetic field can turn the second-order
acceleration more rapid than the first-order one.  This, however,
requires quite ideal turbulence conditions with particle-scattering
waves moving in opposite directions over a sufficiently long
lengthscale.  The Alfv\'en speed is plotted in
Fig.~\ref{fig:alfven_pe} for the same plasma parameters as
in Fig.~\ref{fig:t_fermi2_pe}.

The problem with turbulence conditions in AGN sources is that they are
very poorly known. This leads to large uncertainties regarding the
processes that depend strongly on the turbulence details -- like
stochastic acceleration. In real relativistic sources its efficiency
is probably less than that given above or simulated by
\citet{VV2005ApJ}, because these assumptions exclude the dissipation
and damping of the turbulence, which would limit the acceleration
region spatially. On the other hand, earlier stochastic acceleration
modelling neglects the possibility for much stronger turbulence with
$\delta B \gtrsim B$ as they assume quasilinear conditions with
$\delta B \ll B$.

%
\subsection{Shear acceleration}
Fermi acceleration without a shock is also possible where ever
scattering centres flow at different speeds, even if the flows are
parallel. An example of this is the longitudinal shear across the jet
radius, where particles can tap into the difference between the fast
core of the jet and the slower (or even motionless) exterior. This
mechanism works both on the sharp contact discontinuity between the
streaming jet plasma and the interstellar plasma, as well as on
smoother flows with gradual shear, with scattering centre speed
varying slowly as a function of the radius of the jet
\citep[]{RiegerDuffy2004}.

The shear acceleration timescale in a relativistic flow, with spatial
scale $R$, depends on the extent of the shear region and
the scattering timescale according to \citep{RiegerEtAl2007}
\begin{equation}\label{eq:t_shear}
  \tau_{\rm shear} \approx 
  \frac{15}{4} \frac{R^2}{\Gamma^4 c^2} \frac{c}{\lambda} \approx 
  \frac{15}{4} \frac{R^2}{\Gamma^4 c r_g}.
\end{equation}
With $\Gamma = 10$, $B = 1$ G, and $R = 10^{15}$ cm we get, for an
electron with $\gamma = 10^4$, timescale of decades and rapid 
acceleration requires that 
\begin{equation}\label{eq:shearconstraint}
\gam^{-1}\bg\blob^2\Gam^{-5}<0.013\tv
\end{equation}

Even for models involving minimal size and fast flows \citep[see the
``needle/jet'' model of][with $R=3 \times 10^{14}$ cm and
$\Gamma=50$]{GhiselliniTavecchio2008}, the acceleration is too slow to
have relevance in minute-timescale flares. Due to the strong
dependence on the spatial scale ($\tau \propto R^2$) and the jet
Lorentz factor ($\tau \propto \Gamma^{-4}$), the shear acceleration
requires extremely fast and narrow jets or otherwise very sharp
transition layers between different parts of the jet.

It is interesting to note, however, that because $\tau_{\rm shear}$ is
inversely proportional to the particle mean free path and, thus,
energy -- in contrast to the first- and second-order acceleration --
it can, in principle, continue to accelerate the highest-energy
particles already energised by the first- or second-order mechanisms.
Furthermore, although with Bohm-type scattering shear acceleration
is too slow, we can estimate the circumstances
where shear acceleration would be significant.  We find that shear
acceleration dominates over the first-order process when
\begin{equation}\label{eq:shear_dominates}
  \frac{\tau_{\rm FI}}{t_{\rm shear}} 
  \sim  \frac{\lambda^2}{R^2} \Gamma^4 > 1,
\end{equation}
which is fulfilled for accelerating electrons whose mean free path
satisfies $\lambda > R / \Gamma^2$. For $\Gamma=10$ this means that
the mean free path has to be greater than one percent of the size of
the flaring region. Naturally the mean free path must be shorter
than the minimum dimension of the blob, leading to \( \Gamma^{-2} <
\frac{\lambda}{R} < 1 \). Depending on the behaviour of the mean free
path, this condition could be reached easily with different scattering
models, but as long as we do not have sufficient data suggesting such
turbulence conditions, we will limit our study to Bohm diffusion
with $\lambda \sim r_g$.

%

\section{Energy loss timescales} \label{sec:t_loss}

\subsection{Radiation losses}

Previous authors have already addressed the radiation loss mechanisms
and timescales, so we only review the basic limitations the losses put
on particle acceleration. A single electron, with Lorentz factor
$\gamma$, moving in a plasma with energy density of the local magnetic
field being $\Umag$ and the energy density in the
radiation field being $\Urad$, undergoes synchrotron and inverse
Comptonisation (IC) losses on timescales
\begin{equation} \label{eq:loss_synchrotron}
  \tau = \frac{3 \, m_e \, c}{4 \, \sigma_{\rm T} \, \gamma \, U},
\end{equation}
where $\sigma_{\rm T}$ is the Thompson cross section and the energy
density $U$ in question is $\Umag$ for synchrotron losses and $\Urad$
for IC losses. 
In the latter case, Equation~(\ref{eq:loss_synchrotron})
assumes that Compton scattering happens in the Thomson
regime. However, this may not be true for high-frequency peaked BL Lac
sources, such as Markarian 501 and PKS 2155-304 discussed in the
introduction. In the Klein-Nishina regime Compton cooling is
suppressed by orders of magnitude, and the timescale given by
Eq.~(\ref{eq:loss_synchrotron}) is radically shorter than what would
be expected for the highest-energy particles in these
sources. Detailed cooling time analysis is beyond the scope of this
acceleration-focused paper, but will be included in subsequent
studies.

It is clear from Eq.~(\ref{eq:loss_synchrotron}) that the timescales
of the synchrotron and IC losses differ only in the energy density
term. For magnetic field the density $\Umag = \frac{B^2}{8\pi}$ is
straightforward to calculate, but the energy density of the radiation
field depends on many things, especially on the source of the photons.
In this paper we assume a ratio of the energy densities and use
that as a free parameter. Because of the otherwise equal form except
for the energy density, the timescales for synchrotron and IC losses
follow a simple relation:
\begin{equation} \label{eq:timescale_ratio}
  \frac{\tau_{\rm synch}}{\tau_{\rm IC}} = \frac{\Urad}{\Umag}, 
\end{equation}
and IC losses dominate over the synchrotron losses when $\Urad / \Umag
> 1$, which is considered to be the case in most of the AGN sources
with high radiation densities especially close to the accretion disc
and BLR clouds. Values of the order of $\Urad / \Umag = 10^2$ or $10^3$ are
typically used in modelling these sources. Questions such as whether
the dominating seed photon field is from an external source or created
by the accelerating particles \citep[see, e.g.][for support for both
alternatives]{BegelmanEtAl2008,CelottiGhisellini2008,GhiselliniTavecchio2008},
or how are the loss timescales changed in the extreme Klein-Nishima
region, are beyond the scope of this paper. Furthermore, we
don't include the effects of an increasing magnetic field, but
note that when taken into account
\citep{SchlickeiserLerche2007} the nonlinear effects due to the change in
the equipartition magnetic field energy density
can lead to synchrotron losses significantly
faster than that expected from standard linear losses. Even the
``nonlinear synchrotron losses'' work on timescales longer than the
fastest acceleration mechanism for the modest $\gamma$ particles.

\subsection{Escape and adiabatic expansion}

For Bohm-type diffusion the electron escape timescale depends on the
size of the acceleration region and the particle gyroradius as follows:
\begin{equation}
  \tau_{\rm esc} \sim \frac{R^2}{r_g c} 
  \sim 2\times 10^{16} \frac{R_{15}^2 \, B_1}{\gamma} \; \rm s
\end{equation}
Calculating the timescales it becomes apparent that escape can play a
significant role in shaping a-few-minute flare only if the spatial
scale is essentially smaller than what is required by the causal
connection argument, which limits the emitting region to sizes smaller
than $\sim 10^{13}$ cm (of the order of 1 AU). Even for spatial scales
that small, only particles with $\gamma \gg 10^9$ are considerably
affected by escape on timescales of the order of minutes, so we can
safely exclude escaping from the dominant restrictions to the particle
acceleration mechanisms in these short flares.

We take also the adiabatic losses due to expansion to be negligibe
within our assumptions. The blob is very likely to be travelling
within a collimated flow, where expansion is expected to be slow and
not to play a major role in the first few minutes of the flare. Even
though the condition and expansion speed is not known for these
sources, a rough estimate \citep[see, e.g.][]{Longair} gives 
\begin{equation}
  \tau_{\rm ad} \sim \frac{R}{v_{\rm s}} 
  \sim 3\times 10^4 \frac{R_{15}}{\beta} \; \rm s
\end{equation}
which, in the shortest parsec or sub-parsec scale jets in blazars,
corresponds to timescales of many hours. For sources like
microquasars, however, where the speed and the degree of
collimation of the jet are probably lower, the adiabatic losses are
more likely to set limits for the particle energy.

\section{Results \& Discussion} \label{sec:results}
\begin{figure}
  \begin{center}
    \includegraphics[width=\linewidth]{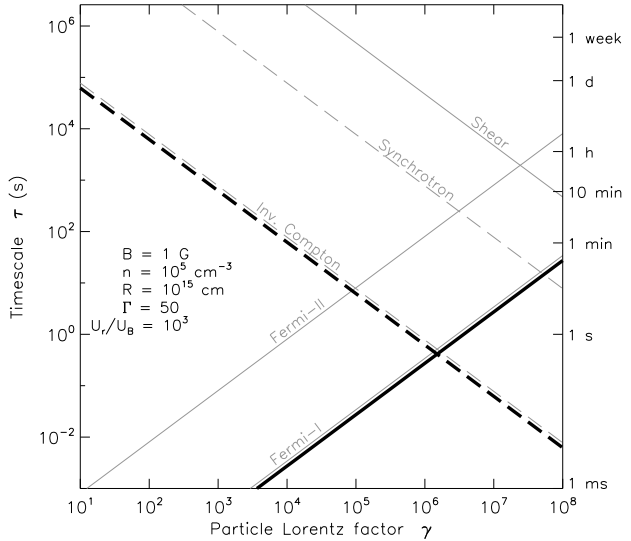}
  \end{center}
  \caption{Calculated timescales in the comoving frame of the
    radiating plasma as a function of the particle Lorentz factor.
    Acceleration timescales (solid lines) are shown for the
    first-order Fermi acceleration (labelled Fermi-I) for electron,
    second-order acceleration (Fermi-II) for an electron in a hydrogen
    plasma, and shear acceleration, and losses (dashed lines) due to
    radiation due to synchrotron and IC; escape losses are too slow to
    appear in this Figure. The thick black lines show the fastest gain
    and loss timescales. Values for the magnetic field $B$, plasma
    number density $n$, bulk Lorentz factor $\Gamma$, and the ratio
    $\Urad / \Umag$ are given in the figure.}    
  \label{fig:timescales}
\end{figure}

We have compared the various acceleration and loss timescales
for various different physical environments by changing the following
parameters: magnetic field intensity, matter density, size of the
region, the jet Lorentz factor, and the ratio of radiation energy
density to that of the magnetic field. In Fig.~\ref{fig:timescales}
the acceleration and loss timescales are shown for a set of
parameters that correspond to the recent models of
\citet{GhiselliniTavecchio2008} and \citet{BegelmanEtAl2008}. The
bold lines correspond to the fastest acceleration (marked with solid
lines) and loss (dashed lines) timescales, and their crossing point
corresponds to the Lorentz factor where gains and losses are equal --
beyond this particles cool instead of accelerating. The most severe
energy loss mechanism is the inverse Compton process, due to $\Urad /
\Umag > 1$. With the aforementioned assumptions, the particle escape
process is orders of magnitude too slow to affect the radiating
electrons significantly within timescales of minutes or hours.

The spatial scale affects the escape and shear acceleration timescales
in the same way: both processes become more important for smaller
sizes, and when considering minute- or even hour-scale variability in
sources with $R \sim 10^{15}$ cm, both are negligible. It is possible,
however, that if the particle scattering mean free path is
significantly larger than the particle gyroradius, both processes
begin to have noticeable effects. 

The first-order process (including the effects of converter
mechanism) accelerates electrons up to $\gamma \sim 10^6$ in just
seconds, so it is fair to say that even in these short flares
acceleration can be considered instantaneous. However, at the same
time the second-order mechanism works on slower but still interesting
timescales. 

The plasma density directly affects only the second-order acceleration
rate through the Alfv\'en-speed dependence: higher density leading to
slower acceleration. Composition of the plasma has the same effect,
because the presence of heavier ions decreases the Alfv\'en speed.

The expected rapid second-order acceleration process raises two 
important questions. Firstly, is the effect of stochastic acceleration seen 
in the observations? And, secondly, if it is not an important process, how 
do we explain this? For the first question, some recent observations
of particle spectra with very hard power-law spectral
indices ($\sigma < 2$, for $N(\gamma) \propto \gamma^{-\sigma}$)
suggest we may see it \citep[]{KatarzynskiEtAl2006,BottcherEtAl2008}.

The latter question, i.e., if we do not see stochastic acceleration
even if the conditions seem perfect for it, suggests basically three
plausible options. 

Turning to the second question, an absence of stochastic
acceleration would raise the following possibilities. Firstly, the
acceleration model could be wrong. However, for the reasons given
later in the discussion, we do not believe this to be the case and
leave this option for other authors. Secondly, it is likely that
we do not fully understand the plasma parameters such as composition
and the magnetic field. Thirdly, the conditions are right, but there
are other factors suppressing the acceleration or turning it off. This
is also a realistic scenario, as we have very little knowledge of the
nature and extent of the turbulence in the acceleration and
radiating regions. A turbulence damping process could shut down
this mechanism. This option is very plausible, because the process
depends heavily on the turbulence in the acceleration regions, and
very little is known about it.  Furthermore, the stochastic
acceleration in relativistic shocks and AGN-related conditions is
still little studied. It is not unlikely that future studies will be
able to set stricter limits for the extent (both spatial and
energetic) of the process, making only a small region
immediately behind the turbulence-creating shock front suitable for
stochastic acceleration.

%
\subsection{Absence of continuous acceleration}

\begin{figure}
  \begin{center}
    \includegraphics[width=\linewidth]{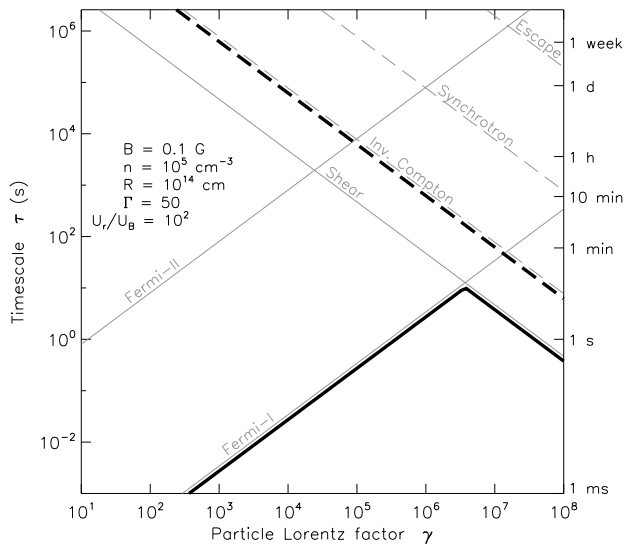}
  \end{center}
  \caption{An example of parameter set which leads to a possible
    dominance of acceleration at all energies: low magnetic field and
    strong shear make the shear acceleration energise high-energy
    particles faster than the IC radiation decreases the energy.}
  \label{fig:only_acceleration}
\end{figure}

It is interesting to note that with the assumption of high-velocity
plasma streaming (whether as a continuous jet or an ejected plasma
``blob'') within an external plasma, there always seems to be suitable
conditions for very rapid particle acceleration. The acceleration can
be ``instantaneous'' first-order acceleration -- feeding the radiating
region constantly with a power law of high-energy particles -- or it
can be  a more gradual process, changing the energy distribution of the
radiating particles on longer timescales.

When looking for a set of parameters that would yield dominating
energy losses -- corresponding to models with instantaneous
acceleration to a power-law spectrum followed by only radiative
cooling and other losses -- we could not find any within 
limits typically considered physical for these sources: 
$B \in [10^{-3},1]$ G; $\Gamma \in [2,50]$; $n \in [1,10^{10}]$ cm$^{-3}$;
$R \in [10^{10},10^{15}]$ cm; $\Urad/\Umag \in [10^{-3},10^{3}]$. 
Acceleration times within these limits 
were always shorter than the loss timescales in the lower energies up
to electron Lorentz factors $\gamma \sim 10^{5-6}$. In fact, in some
cases the ``inverse'' energy dependence of the shear acceleration led
to domination of acceleration on all energies -- an example of this is
shown in Fig.~\ref{fig:only_acceleration}. Although we don't expect
this to be a typical case in real sources (for reasons discussed
earlier) it underlines the problem of getting only cooling without
having acceleration at the same time.

%
\subsection{Time delays}

Interesting constraints on acceleration mechanisms arise from 
observations of a TeV flare of Markarian 501 on
July 9th, 2005. Here the hard gamma-rays (2.7 TeV) were observed to
lag behind the soft ones (190 GeV) by roughly four minutes.  Similar
``hard lags'' have been observed earlier also in the X-ray domain
\citep[e.g.,][for Markarian 401]{BrinkmannEtAl2005}, but no definite
explanation has been found yet.  The ``soft lags'' -- where lower-energy
radiation peaks later than the higher-energy one -- are easier to explain
by high-energy radiating particle cooling and radiating on lower
and lower frequencies, but, intuitively, these hard lags would require
particles not cooled but \emph{heated} or accelerated during the flare.
 
The observed delay has been proposed as being sue to the radiating blob
accelerating during the flare by \citet{BednarekWagner2008}, but even
in their model the particles would only undergo cooling without any
ongoing acceleration in or around the high-$\Gamma$ plasma
flow.  \citet{MactichiadisMoraitis2008}, however, showed that the
observed features can be explained with a very simple and physically
plausible modification, namely allowing the particles to accelerate
gradually.  In their model the acceleration time is of the order of
hours and the electrons reach energies corresponding to $\gamma \sim
10^6$. The timescale may be too long for the first-order mechanism, but,
interestingly, matches well to the stochastic acceleration timescale
corresponding to their parameters.

Furthermore, similar slow energisation was quantitatively predicted by
\citet{VV2005ApJ}, whose simulations of stochastic acceleration in
relativistic shocks showed a gradual shift of the whole particle
spectrum to higher energies as the particles were accelerated behind a
shock front. Hadronic process could also be relevant within this model.

%
\subsection{Generation of turbulence}

Finally, one must also remember that as the accelerating particles need
electromagnetic turbulence for scattering and energy gain, it is
obvious that suitable turbulence either has to exist in the pre-shock
plasma or be generated at the shock by the particles themselves. All
models with instantaneous or constant first-order Fermi acceleration
require the acceleration timescale to be significantly less than the
flaring time, and the same limit is set also for the turbulence
generation. Recent numerical work has addressed the problem of
turbulence generation, and although the problem is still far from
being solved, studies such as that by \citet{RevilleEtAl2006}, suggest
that especially nonlinear turbulence generation by the particles
themselves \citep[found by][]{Bell2004} could indeed be efficient
enough to provide sufficient turbulence for fast acceleration.

\section{Conclusions}

We have estimated the timescales of different particle-acceleration 
mechanisms in the context of minute-scale TeV blazar flares. We have 
excluded both the neutron-based converter mechanism and 
shear acceleration from the dominating processes in these sources on
observed timescales of the order of minutes. Furthermore, the
first-order Fermi acceleration of protons is likely to be too slow to
appear ``instantaneous''. Instead, the accelerated proton distribution
is expected to continue to be energised during the flare and could
have special significance in hard-lag sources.

Various simultaneously active acceleration mechanisms working on
different timescales suggest that in these sources one could expect to
find slower, gradual energisation in addition to instantaneous -- or a
brief initial period of -- acceleration. This speaks in favour of
models with ongoing acceleration, instead of models with instantaneous
injection of a fully-developed power-law spectrum which then only
undergoes cooling.  However, since many models that only include
cooling can reproduce the observations, their results combined with
timescale analysis could be helpful in studying the source parameters.

We emphasise that the current model is very simplified. The
results neglect, in particular, the Klein-Nishina effects in the
energy losses of the highest-$\gamma$ particles. Also, the magnetic
field is assumed to be constant throughout the flare, although
generation of turbulence and the subsequent increase in the magnetic
field strength can decrease the synchrotron loss timescales
significantly \citep{SchlickeiserLerche2007}. Furthermore, because the
acceleration efficiency depends on the scattering mean free path, for
different scattering models the acceleration timescales could also
change.  Furthermore, other omitted
turbulence-affecting effects can decrease the acceleration
timescales, for example by increasing the acceleration efficiency. An
example is the case of turbulence transmission leading to an increased
scattering-centre compression ratio (see Vainio et al.,
\citeyear{VVS2003,VVS2005}, and Tammi \& Vainio \citeyear{TV2006}), 
and enhanced first-order acceleration at the shock in
conditions similar to AGN or microquasar jets
\citep{Tammi2008}. However, even in its current state, especially in
the lower-energy part of the particle spectrum ($\gamma \lesssim
10^4$) the present analysis still provides a useful tool for
estimating the source properties.

Finally, we note that since the acceleration efficiency depends
on the scattering mean free path differently for each of the
processes, detailed modelling could provide additional tests when we
learn more about the intrinsic properties of the plasma and turbulence
in these sources.

\section*{Acknowledgments}

J.~T.~was funded by the Science Foundation Ireland grant 05/RFP/PHY0055.

%
%

\bsp

\label{lastpage}

\end{document}